# Flow rate measurement in stacks with cyclonic flow – Error estimations using CFD modelling

J. Geršl[a,*], S. Knotek[a], Z. Belligoli[b], R. P. Dwight[b], R. A. Robinson[c], M. D. Coleman[c]

[a]*Czech Metrology Institute, Okružní 31, 63800 Brno, Czech Republic*
[b]*TU Delft, Mekelweg 2, 2628 CD Delft, The Netherlands*
[c]*National Physical Laboratory, Hampton Road, Teddington, Middlesex, TW11 0LW, UK*
*Corresponding author, E-mail address: jgersl@cmi.cz*

**Abstract**

Two methods of flow measurement in stacks are investigated to determine their errors in presence of cyclonic flow. One method – based on velocity measurements with a Pitot tube in a grid of points – is the standard reference method according to EN ISO 16911-1. The second method – ultrasonic flow measurement – is often used as the automated measurement system in stacks according to EN ISO 16911-2. Several typical stack configurations are considered and the flow field in the stacks is obtained using validated computational fluid dynamics (CFD) modelling with OpenFoam software. We show that possible errors of the standard reference method due to the cyclonic flow are significant compared to the requirements of the EU's Emissions Trading System. For the ultrasonic flow meter we compare various configurations (number, orientation, position) of the ultrasound beams and we demonstrate the flow profile pre-investigation by CFD as prescribed in section 8.3 of EN ISO 16911-2.

## 1. Introduction

The measurement of flow rate in stacks is an important component for determining the annualised mass emissions of pollutants released to the atmosphere. Current regulations, e.g. the EU's Emissions Trading Scheme (EU ETS) [1], are introducing lower emission limit values and new challenging uncertainty requirements for emission measurement methods. For example the Monitoring and Reporting Regulations [2] of the EU's ETS specify a maximum permissible uncertainty of 2.5 % for annual emissions from large emission sources (>500000 tonnes of $CO_2$ per annum). This gives also the uncertainty limit for the flow rate measurement. In order to meet the uncertainty requirements of the EU's ETS CEN published EN ISO 16911-1 [3] superseding ISO 10780 [5], which was not able to meet such requirements [6].
One of the most common methods used as the standard reference method (SRM) for flow rate measurement in stacks is the measurement of speed of the gas with S-type Pitot tubes in a grid of points inside the stack [3-5]. Accurate measurement using this method is challenging especially when non-axial velocity components are present since the indication of the S-type Pitot tube is sensitive to the orientation of the Pitot head with respect to the gas flow direction [11-14]. The standards [5, 9] describe how to identify and measure the tangential velocity component (swirl) in a stack with the Pitot tube and [3] then introduces a correction for the swirl that is applied if the swirl angle (yaw angle) exceeds 15°. However, the radial velocity components that also influence the error of the Pitot tubes are not considered by the standard, as neither are swirl angles below 15° (see [6] for a summary). Moreover, if an asymmetric velocity profile occurs, it can rotate with height in the stack due to the swirl and special attention must be paid to the orientation and height of the measurement ports when we want to minimize the flow measurement error. The standard [3] was validated via field trials [7], however, the trials were carried out at plants with no significant cyclonic flow so this part of the standard has not been validated [6], which gives a further motivation for investigating the effects of swirl.

Another common method used as automated measuring system (AMS) in stacks [8] is the ultrasonic flow technique. The indication of ultrasonic flow meter is directly related to the average tangent flow velocity along the ultrasonic beam (see section 2.4 of this paper). Therefore, to determine the flow rate from the ultrasonic measurement assumptions about the complete flow field (axial velocity profile and non-axial velocity components) must be made and these can lead to errors if they deviate from reality. For review of the ultrasonic flow measurement including the velocity profile effects see e.g. [15]. For recent works on CFD modelling of ultrasonic flow meters see e.g. [16, 17]. The standard [8] describes how the velocity profiles in stacks should be pre-investigated before the measurement device is installed and gives recommendations about the measurement paths used for given velocity profile characteristics.

The pre-investigation can be done by measurement (section 8.2 of [8]) or if it is not possible, e.g. if the plant is not yet built or if the duct configuration is too complicated for the prescribed conditions for the measurement to be fulfilled, the measurement may be replaced by computational fluid dynamics (CFD) (section 8.3 of [8]). The CFD modelling can be used to determine the expected flow profile changes as a function of plant operation conditions, to assist the selection of the AMS type and to determine the optimal position of the AMS.

In this paper we use a CFD model to determine flow fields in a vertical circular stack with a perpendicular supply pipe of three different configurations generating different swirls, namely a) straight pipe, b) single 90° elbow and c) double 90° out of plane elbow. For each of the configurations three different flow rates are considered corresponding to inlet velocities of 3 m/s, 10 m/s and 30 m/s. We use OpenFOAM software validated by experimental data to model the flow. The obtained velocity fields are then used to test the above flow measurement methods (S-type Pitot tube and ultrasonic meter) for errors.

For the measurement with the S-type Pitot tube in a grid of points according to [3] we calculate the flow rate error due to the finite grid density and we show how this error depends on height in the stack where the shape of the flow profile as well as the orientation of the flow profile maximum velocity with respect to the measurement grid are changing. Then we calculate the flow rate error due to the nonzero angle between the velocity vector and the measurement direction of the meter and we show the improvement of the error if the correction for swirl according to [3] is applied. Part of the results for the Pitot tubes has been published in [18]. We include these results also in this paper for completeness.

For the ultrasonic flow measurement we compare errors for several measurement setups – one and two path configurations with various orientations of the ultrasonic beams. We determine the flow profile characteristics defined in the pre-investigation procedure of the standard [8] and in the comparison we include the measurement setups that are recommended by [8] for these actual flow profile parameters.

**2. Flow measurement in stacks**

In this section we review the principles of flow measurement in stacks using S-type Pitot tubes with grid sampling according to [3] and with the ultrasonic flow meters. We summarise some basic formulas that are later used for calculation of the meter indication for velocity fields simulated by the CFD.

*2.1 Flow measurement using velocities in a grid*

This method uses a grid of points in a plane perpendicular to the stack axis where each point "covers" the same area. The exact point distribution and the minimal number of points for a given stack diameter are prescribed in [4]. Typically the points are distributed along two lines in several circles with radius given by the same area requirement. A point distribution with two lines and three circles is shown in Figure 1.

We denote $v_{z\alpha}$ the velocity component along the stack axis (*z*-axis) measured in the point $\alpha$ with $\alpha = 1, ..., N$ where $N$ is the total number of measurement points. The measured flow rate is then determined as

$$Q_M = \frac{A}{N} \sum_{\alpha=1}^{N} v_{z\alpha} \qquad (1)$$

where $A$ is the area of the stack cross section.

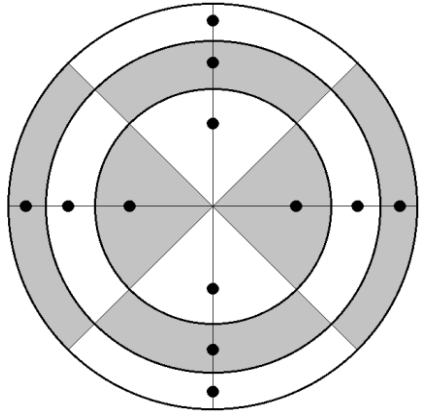

**Fig. 1.** Flow rate measurement grid – all the grey and white fields have the same area. The position of the points in the fields is defined such that a circle going through the point cuts the field into two parts with the same area.

*2.2 Angle dependence of the S-type Pitot tubes*

Pressure difference across a Pitot tube depends on the mutual orientation of the gas velocity vector and the Pitot tube which is defined by the yaw and pitch angles. The orientation of the S-type Pitot tube is given by its tangent vector $\vec{t}$ and measurement direction vector $\vec{z}$ as shown in the Figure 2.

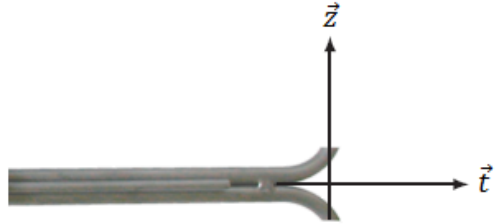

**Fig. 2.** Vectors defining the orientation of the S-type Pitot tube.

The pitch angle is then defined as the angle between the velocity vector and the vector $\vec{t}$ minus 90°. The yaw angle is defined as the angle between the vector $\vec{z}$ and projection of the velocity vector to the normal plane of $\vec{t}$. The yaw angle is positive when the velocity vector is pointing to the $\vec{z} \times \vec{t}$ direction.

During a calibration in a wind tunnel the Pitot tube can be turned with respect to the main flow direction and a dependence of the calibration factor $C(\beta,\gamma)$ on the yaw angle $\beta$ and the pitch angle $\gamma$ can be determined from the equation

$$v = C(\beta,\gamma)\sqrt{\frac{2\Delta p(\beta,\gamma)}{\rho}} \qquad (2)$$

where $v$ is the reference velocity in the wind tunnel, $\rho$ is the air density in the wind tunnel and $\Delta p(\beta,\gamma)$ is the differential pressure across the Pitot tube turned by the angles $\beta$ and $\gamma$. The velocity component in the $\vec{z}$ direction is then given as

$$\begin{aligned} v_z &= v.\cos\beta.\cos\gamma \\ &= C(\beta,\gamma).\cos\beta.\cos\gamma\sqrt{\frac{2\Delta p(\beta,\gamma)}{\rho}} \end{aligned} \qquad (3)$$

If the S-type Pitot tube is installed in a stack and well aligned such that the $\vec{z}$ vector of the tube lies along the stack axis the calibration factor $C(\beta,\gamma).\cos\beta.\cos\gamma$ should be used to determine the $v_{z\alpha}$ value in case of a swirling flow in the stack. If the swirl is neglected and the calibration factor $C(0,0)$ is used instead the resulting relative percentage error of the velocity measurement is given as

$$E = \left(\frac{C(0,0)}{C(\beta,\gamma).\cos\beta.\cos\gamma} - 1\right).100 \tag{4}$$

The yaw and pitch angle dependence of the S-type Pitot tubes has been investigated by several authors in the past [11-14]. For the purpose of this paper we use data published in [11] in Figure 4-12 of this reference. We restrict the range of the yaw angles to (-25°, 25°) and the pitch angles to (-10°, 10°) and fit the experimental data for the error (4) by second order polynomials. The resulting formula is

$$E = (a_1\gamma^2 + a_2\gamma + a_3)\beta^2 + c_1\gamma^2 + c_2\gamma \tag{5}$$

where $E$ is the error in percents, $\gamma$ and $\beta$ are the pitch and yaw angles in degrees and $a_1 = 1.4 \times 10^{-5}$, $a_2 = 4.5 \times 10^{-4}$, $a_3 = 0.024$, $c_1 = 5.9 \times 10^{-3}$ and $c_2 = 0.13$. The formula (5) is specific for the instrument reported in [11]. Different instruments can have significantly different errors. E.g. one of the Pitot tubes reported in [12] has errors which are more than double that of the instrument considered in this paper.

*2.3 Errors of the flow measurement with Pitot tubes*

The flow rate determined by the grid "integration" (1), where we suppose that the measured velocities $v_{z\alpha}$ are exactly the z-components of the velocity, is denoted as $Q_{Mz}$ in the following text. Therefore, the error of the flow rate $Q_{Mz}$ comes only from the approximate integration and not from the inaccurate velocity measurement in particular points.

On the other hand, the flow rate determined by (1), where we suppose that the measured velocities $v_{z\alpha}$ have an error (4) due to the neglected yaw and pitch angles, is denoted as $Q_{Mi}$. In this case the error of $Q_{Mi}$ comes both from the approximate integration and from the velocity measurement error.

When the swirl is significant the standard [3] describes how the velocity measurement should be corrected in section 9.3.5. The yaw angle of the cyclonic flow can be determined by rotating the S-type Pitot tube around its axis and finding a position with zero indication of the differential pressure as described in Annex C of [5] or in section 11.4 of [9]. In that case the velocity vector is normal to the $\vec{z}$ vector of the Pitot tube (see Figure 2) and the angle between the $\vec{z}$ vector and horizontal plane is the yaw angle of the flow at the particular point. If the angle is larger than 15° the standard [3] prescribes a correction procedure for the measured velocity. The velocity $v_{meas}$ is measured along the actual flow direction, i.e. with the Pitot tube turned by the previously measured yaw angle $\beta_{meas}$. The corrected axial velocity component $v_C$ is determined according to [3] from

$$v_C = v_{meas} \cos\beta_{meas} \tag{6}$$

This procedure with S-type Pitot tube still does not take the pitch angle into account and therefore the velocity $v_C$ is still biased. The relative error of $v_C$ with respect to the real axial velocity is given by the formula (4) with $\beta = 0$.

When the flow rate is calculated based on velocities that are corrected according to section 9.3.5 of [3] for yaw angles exceeding 15° we refer to this flow rate as corrected $Q_{Mi}$.

If we denote $Q$ the real flow rate in the stack we can define the relative percent errors as

$$E_z = \frac{Q_{Mz} - Q}{Q}.100 \tag{7}$$

$$E_i = \frac{Q_{Mi} - Q}{Q}.100 \tag{8}$$

*2.4 Ultrasonic flow measurement*

The setup of ultrasonic flow meters is shown in Figure 3. The ultrasonic transducers are installed at the stack wall – one of them upstream and another one downstream.

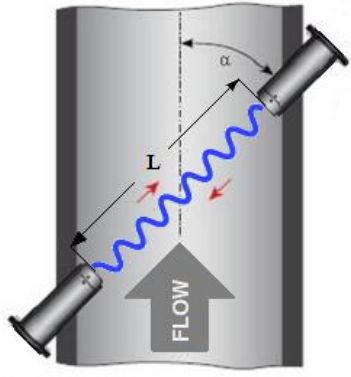

**Fig. 3.** Ultrasonic flow measurement in a stack.

We denote the time of ultrasonic signal propagation from the upstream to the downstream transducer as $t_d$ and the time of propagation from the downstream to the upstream transducer as $t_u$. Hence

$$t_d = \int_0^L \frac{dl}{c + \boldsymbol{v}.\boldsymbol{e}}, \qquad t_u = \int_0^L \frac{dl}{c - \boldsymbol{v}.\boldsymbol{e}} \tag{9}$$

where $\boldsymbol{e}$ is unit tangent to the ultrasonic beam pointing from the upstream to the downstream transducer, i.e. $\boldsymbol{v}.\boldsymbol{e}$ is the gas velocity component tangent to the beam, and $L$ is the distance between the transducers ($l$ is length coordinate along the beam with $l = 0$ at the upstream transducer). The time $t_d$ is shorter than $t_u$ since the signal travelling downstream is going with the gas flow whereas the signal travelling upstream is going against it. Therefore the time difference is related to the flow velocity. If the flow velocity $\boldsymbol{v}$ is small enough compared to the speed of sound (in the gas being at rest) $c$ we have

$$t_u - t_d \approx \frac{2}{c^2} \int_0^L \boldsymbol{v}.\boldsymbol{e}\, dl \tag{10}$$

Therefore, the measured time difference is directly related to an approximation of the true average tangent gas velocity along the beam. In ultrasonic flow metering the times $t_d$ and $t_u$ are measured and the flow rate indication of the meter is given as [15]

$$Q_M = \frac{AL}{2\cos\alpha} \frac{t_u - t_d}{t_u t_d} \tag{11}$$

where, $A$ is the area of the stack cross section and $\alpha$ is an inclination angle of the beam from the stack axis (see Figure 3). In terms of the velocity field we can express the measured flow rate (11) using (9) and (10) as

$$Q_M \approx \frac{A}{\cos\alpha} \frac{1}{L} \int_0^L \boldsymbol{v}.\boldsymbol{e}\, dl \tag{12}$$

The formulas (11) or (12) give the real flow rate (zero error of the indication) if the flow velocity is axial and constant across the pipe. In real situations, however, this is not satisfied due to the boundary layer near the walls or due to upstream disturbances. In the simplest case of a fully developed turbulent flow, which is axial and the shape of its flow profile can be predicted, the non-uniformity of velocity can be taken into account by a correction factor to the formula (11) (this factor converts the line average of the velocity in (12) to the cross section average needed for the flow rate). However, if upstream disturbances are present the flow profile can be even more distorted and non-axial velocity components can occur. This is what causes the flow meter errors since there is not a way to compute a general correction factor because of the large variety of different possible flow fields. In that case the number of paths of the flow meter can be increased and the flow velocity can be evaluated by a weighted average of the velocities measured by each path [15].

## 3. CFD modelling of flow in stacks

In this section we describe the modelled physical situations and the CFD model used to obtain the velocity fields. Next we show the results of the CFD modelling – the flow velocity fields for the specified stack geometries and flow rates – and we calculate their characteristics as prescribed in [8].

*3.1 The studied cases*

In our investigation we consider a stack with circular cross section of diameter 1.5 m and length of 18 m from the bottom to the outlet. A supply pipe is connected to the stack with its axis at the height of 2.25 m from the bottom. The supply pipe can have various shapes but its cross section is always circular with 1.5 m diameter – same as the stack. The connection is a 90° T-junction. Three configurations of the supply pipe are considered, each of them generating a different swirl pattern: (a) a straight pipe; length of the pipe is 5 m from the inlet to the stack wall (Figure 4a), (b) a pipe with a single 90° elbow; the radius of the elbow is 1.5 m; the length of the straight parts upstream and downstream of the elbow is 3 m (Figure 4b) and (c) a pipe with double 90° out of plain elbow; the radius of both elbows is 1.5 m; the length of the straight parts upstream, downstream and in between the elbows is 3 m (Figure 4c).

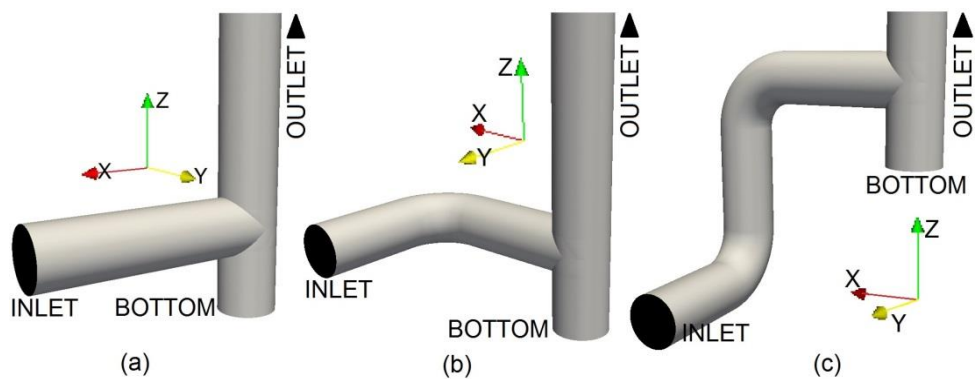

**Fig. 4.** (a) the straight supply pipe, (b) the supply pipe with single 90° elbow and (c) the supply pipe with double 90° out of plain elbow. The bottom of the stack is blind. The direction in which the outlet occurs is marked by arrow.

For each of the three configurations of the supply pipe we study three cases with different inlet velocity values. The considered velocity fields at the inlet to the supply pipe are homogeneous with magnitude of 3 m/s, 10 m/s and 30 m/s. From the outlet of the stack the gas is released to the atmosphere and the corresponding boundary conditions are specified (see the section 3.2.2). The bottom of the stack is closed and treated as a wall. A gas with kinematic viscosity of $\nu = 15$ mm$^2$s$^{-1}$ is considered in the stack which corresponds to air at a temperature of around 20 °C. Thermal effects influencing the flow are not taken into account.

*3.2 The CFD model*

The OpenFOAM software was used for the CFD modelling. OpenFOAM is a free open-source package containing tools for all steps of the CFD procedure – geometry and mesh generator, solvers and post-processing tools for data analysis and visualisation. The computations were performed on the cluster of CMI (544 processor cores, 6 TB RAM) using a parallel run on 16 cores. The computation time needed for one case was between 34 and 90 hours.

*3.2.1 The mesh*

The computation mesh is a structured mesh created by the *blockMesh* tool of OpenFOAM. Several mesh densities have been tested to achieve a mesh-converged solution. The parameters of the final meshes are summarised in Table 1.

| supply pipe | number of cells | y+ for 10 m/s | cell size at the wall | cell size in the centre |
|---|---|---|---|---|
| straight pipe | 16M | 20 | 0.6 mm | 1.8 cm |
| single elbow | 30M | 20 | 0.6 mm | 1.5 cm |
| double elbow | 37M | 20 | 0.6 mm | 1.5 cm |

**Table 1** Parameters of the meshes

The final mesh in the stack cross section for the cases with elbows is shown in Figure 5a. Detail of the T-junction with the structured mesh created in *blockMesh* is shown in Figure 5b.

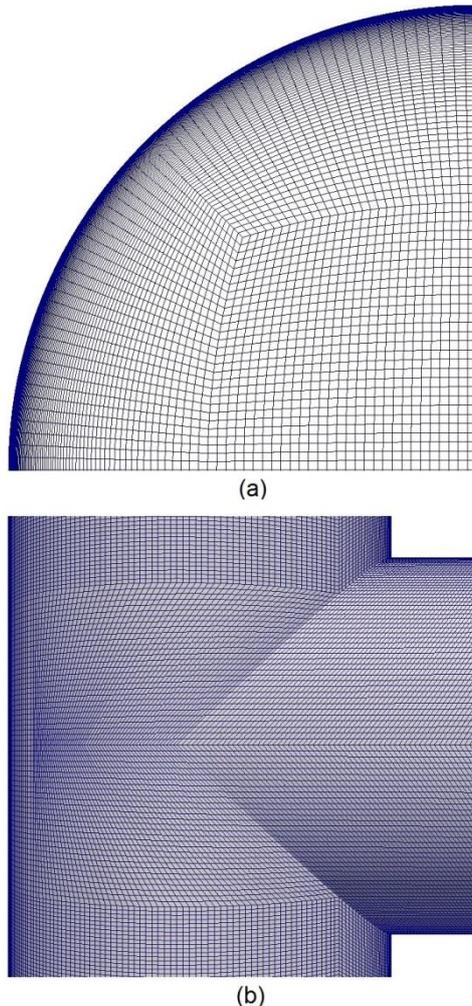

**Fig. 5.** (a) Mesh in the pipe cross section. (b) Mesh in the T-junction created in *blockMesh*.

*3.2.2 The model settings and boundary conditions*

The flow is computed as a stationary incompressible viscous turbulent flow without thermal effects using the *simpleFoam* solver of OpenFOAM. Several types of RANS turbulence models were examined and the CFD model was validated by comparing the results to experimental data published in literature [19] before the simulations for the stacks were started (see the Appendix for details). In the end, OpenFOAM's *kEpsilon* turbulence model was selected for the simulations.

At the inlet of the supply pipe a constant normal velocity field is prescribed with magnitude of 3 m/s, 10 m/s or 30 m/s; at walls the no-slip boundary condition is prescribed and at the outlet from the stack a zero normal gradient is prescribed for the velocity. The pressure is fixed at the outlet and zero normal gradient is prescribed at inlet and at the wall. The values of the turbulence fields $k$ and $\varepsilon$ are fixed at the inlet and correspond to a turbulence intensity of 5 %. Wall-functions are applied at the walls and at the outlet the zero normal gradient condition is prescribed.

The results for the supply pipes with elbows presented below are obtained from solutions with residuals of the velocity and turbulence fields converged to the order of $10^{-6}$ and residuals of the pressure converged to the order of $10^{-4}$. For the straight supply pipe the residuals are a little bit higher – $10^{-5}$-$10^{-6}$ for the velocity and turbulence fields and $10^{-3}$ for the pressure.

*3.3 Computed velocity fields in the stacks*

Figure 6 shows the swirl structures in the stack generated by the three configurations of the supply pipe: (a) the straight pipe generates two counter-rotating swirls, (b) the single elbow pipe generates a single clockwise (view from the stack outlet) swirl, (c) the double elbow pipe generates a single counter-clockwise (view from the stack outlet) swirl. The plots are taken at the height of 7 m above the stack bottom for the cases with inlet velocity of 10 m/s. For other velocities the character of the swirl remains the same.

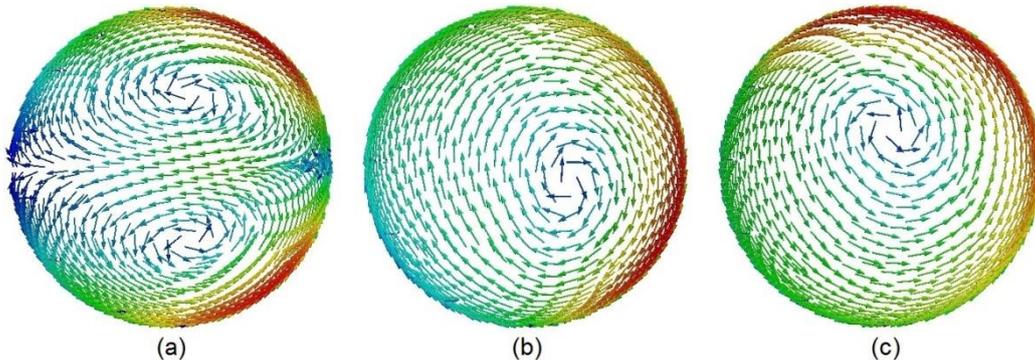

**Fig. 6.** The swirl structures appearing in the stack with (a) the straight supply pipe, (b) the supply pipe with single elbow and (c) the supply pipe with double elbow.

In Figure 7 we show the computed velocity profiles of the axial velocity $v_z$ at several cross-planes along the stack at heights of 9, 11, 13, 15 and 17 m above the bottom for the three configurations of the supply pipe with inlet velocity 3 m/s. In all cases an asymmetric profile occurs that is created behind the T-shape junction of the supply pipe since most of the gas flows up near to the wall opposite to the inlet. We can then look how the position of the maximum of the velocity profile develops with height, e.g. by observing the angle between the *x*-axis and a line connecting the maximum with the centre of the stack. In the case of the straight supply pipe (a) the maximum stays at a 180° angle; for the single elbow case (b) we see a clockwise rotation of the velocity profile corresponding to the clockwise swirl in the stack and for the double elbow case (c) we see a counter-clockwise rotation of the velocity profile corresponding to the counter-clockwise swirl in the stack. The dependence of the angular position of the maximal axial velocity on the height is shown in Figure 8 for all the cases. We see that the rotation rate of the axial velocity maximum is approx. 90° per 8.0 m of the stack height in the case of the single elbow at 3 m/s, and it is approx. 90° per 4.4 m of the stack height in the case of the double elbow at 3 m/s. In both cases the stack height needed for the 90° turning is increasing with the flow rate.

In Figures 9 and 10 we show the distribution of the yaw and pitch angles for various configurations of the supply pipe. The sign convention for the angles in these figures corresponds to a Pitot tube with the $\vec{t}$ vector pointing towards the stack axis. We show only the result for inlet velocity of 3 m/s where the swirl is the most significant. With increasing inlet velocity the maximal yaw and pitch angles are decreasing in all cases and the trends remain unchanged. In [3] a correction for a cyclonic flow is prescribed that is applied if the yaw angle exceeds 15° in a grid point. In Figure 9, we can see the regions of the absolute value of the yaw angle exceeding 15°. They occur for the supply pipes with single and double elbow and they have a red or black colour in the scale. We see that the yaw and pitch angle profiles rotate with the changing height in a similar way as the axial velocity profiles.

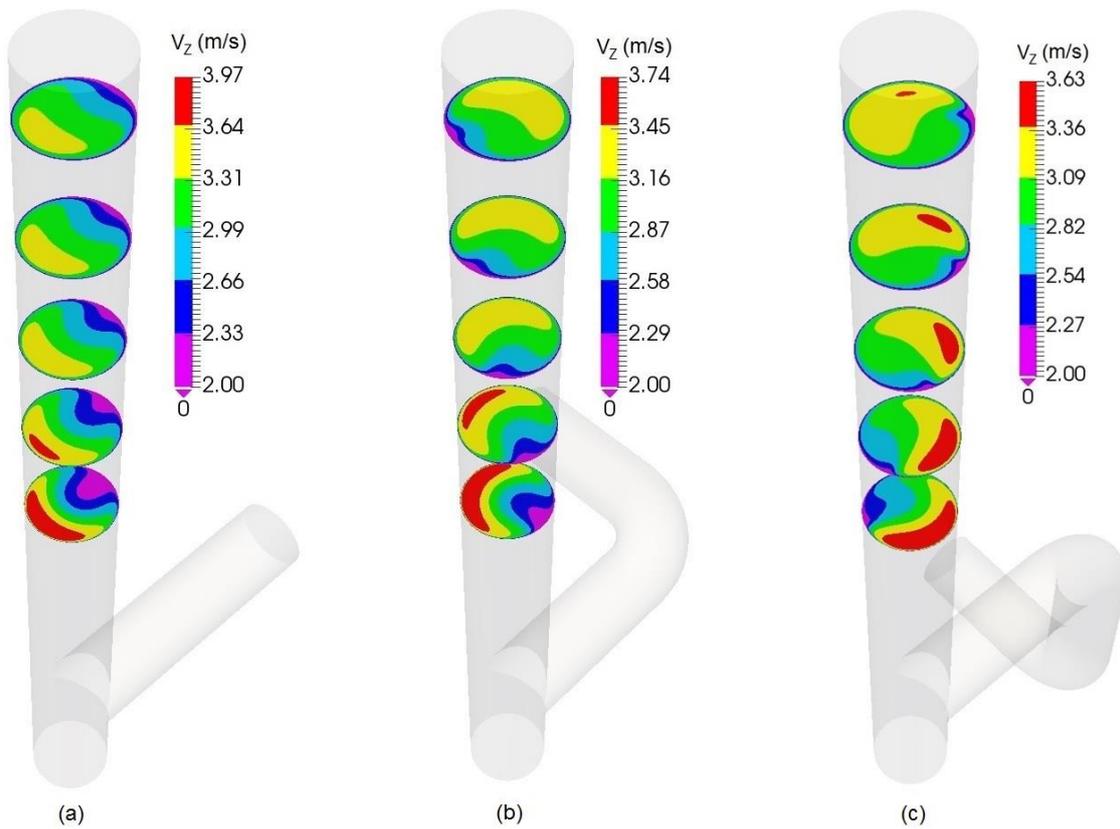

**Fig. 7.** Distribution of axial velocity component in cuts along the stack for (a) straight supply pipe, (b) single elbow supply pipe and (c) double elbow supply pipe with inlet velocity of 3 m/s. The lowest cut is at height of 9 m above the bottom of the stack and the interval between the cuts is 2 m.

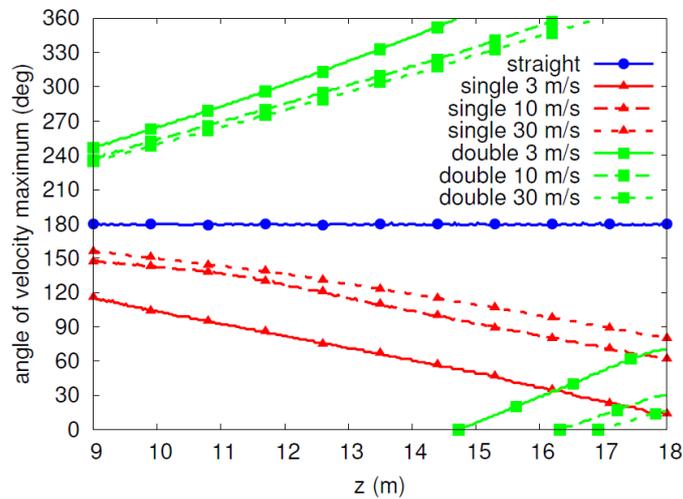

**Fig. 8.** Angular position of the maximal axial velocity as a function of height.

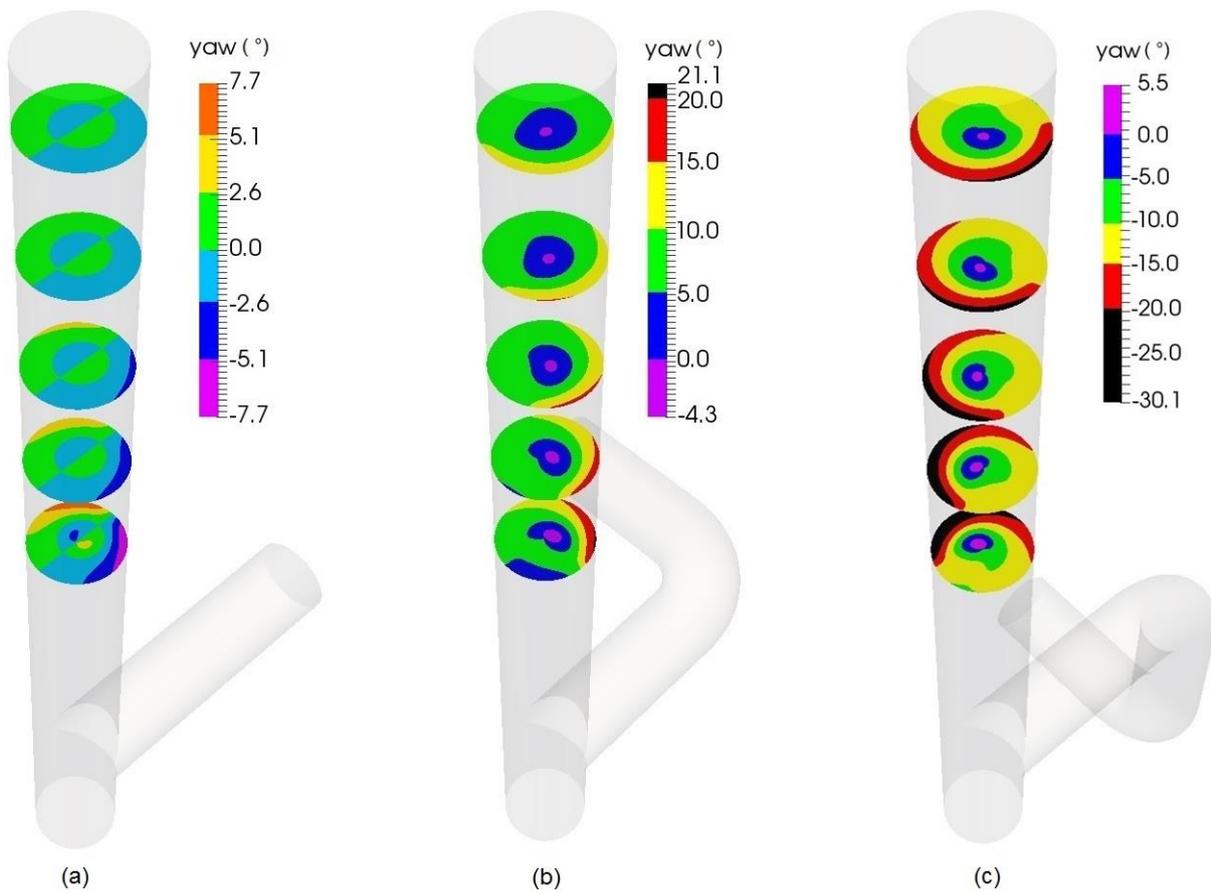

**Fig. 9.** Distribution of the yaw angle for an inlet velocity of 3 m/s.

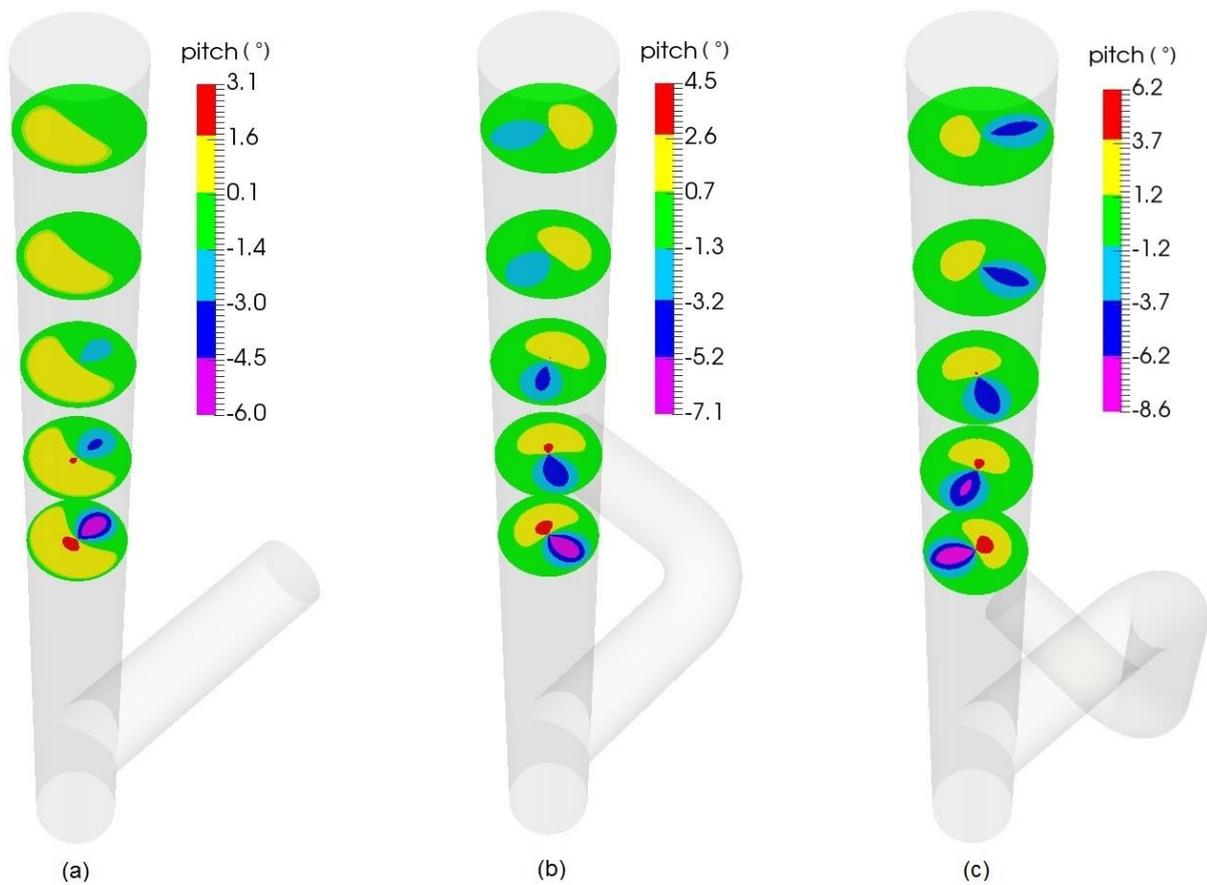

**Fig. 10.** Distribution of the pitch angle for an inlet velocity of 3 m/s.

*3.4 Flow profile characteristics*

The standard [8] defines certain flow profile characteristics in Annex B, namely the crest factor and skewness. Another parameter required for the flow profile pre-investigation is reproducibility as defined in Annex F. These characteristics are then used to select the appropriate flow measurement method and its installation. The flow profile characteristics in the standard are related to flow profiles on sampling lines where the velocities can be measured. Here we use the advantage of the CFD model, which gives us the velocity field in the complete geometry and we define the flow profile characteristics using a complete velocity field in planes given by cuts of the stack at certain heights instead of using only the velocities on the sampling lines. We define the crest factor as the ratio

$$cr = \frac{v_{z,max}}{v_{avg}} \qquad (13)$$

with $v_{z,max}$ being the maximum of the axial velocity component in the given plane and $v_{avg}$ being the average axial velocity in the plane. Next we define the skewness as a ratio

$$sk = \frac{v_{avg+}}{v_{avg-}} \qquad (14)$$

where, $v_{avg+}$ and $v_{avg-}$ are average velocities in two half-planes (half-circles) divided by a line that is going through the centre of the stack and is perpendicular to a line connecting the velocity maximum with the centre of the stack. The value $v_{avg+}$ is for the half-plane containing the velocity maximum. With this definition of half-planes the maximal skewness is determined approximately.

The values of the crest factor and the skewness of the (2D) velocity profiles in various heights are shown in Figure 11 for all the investigated cases.

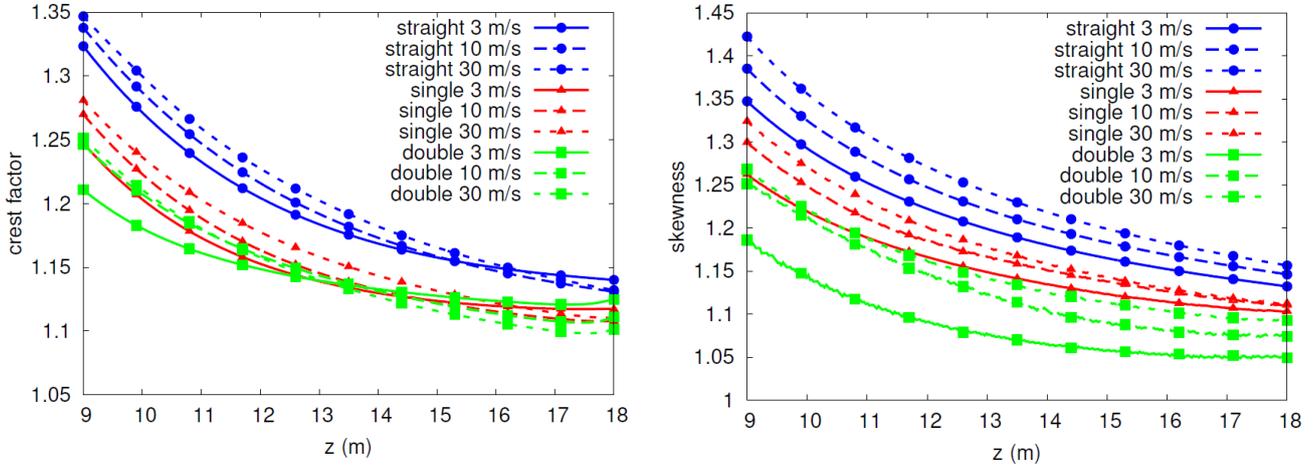

**Fig. 11.** Crest factor and skewness as a function of height.

The pre-investigation procedure of [8] prescribes to measure the flow profile for two operating conditions – "a) one where the flow profile is expected to be most uniform, i.e. close to the highest possible flow rate and with the least possible obstruction to the flow path" and "b) one where the flow rate is so low, that it is not occurring lower more than 10 % of the plant's normal operation time or the point of minimum stable operation, combined with the maximum obstruction to the flow path, e.g. closure of dampers or regulation of the fan blowers." In other words, the purpose is to test the widest possible range of different flow profiles that can occur during the plant operation and to compare them. The comparison is done by calculating the reproducibility of the flow profile according to Annex F of [8]. Here the reproducibility is defined as

$$R = t_{0.95(n-1)} \sqrt{\frac{\sum_{i=1}^{n}(x_{1i} - x_{2i})^2}{2n}} \qquad (15)$$

where, $x_{1i}$ is the velocity in the $i$-th measurement point normalised by the average velocity for the high flow conditions (a), $x_{2i}$ is the velocity in the $i$-th measurement point normalised by the average velocity for the low flow conditions (b), $n$ is the number of measurement points and $t_{0.95(n-1)}$ is the two-sided Student $t$-factor at a confidence level of 0.95 with $n-1$ degrees of freedom, as given in Annex E of [8].

In the standard [8] the reproducibility is evaluated for flow profiles along the measurement lines only. In this paper we use the advantage of knowing the complete velocity field from the CFD model and we calculate the reproducibility of flow profiles in complete 2D cuts of the stack at various heights, i.e. we use the formula (15) applied to a grid of points as defined by [4] covering all the 2D cuts with high density. The grid we use has 180 measurement lines (1° angular step) with 600 points each, i.e. $n = 108\,000$. We expect that a further increase in the grid density will not change the result for reproducibility significantly. The reproducibility is calculated by comparing the flow profiles for inlet velocities of 3 m/s and 30 m/s for each of the three geometries. The result is shown in the Figure 12.

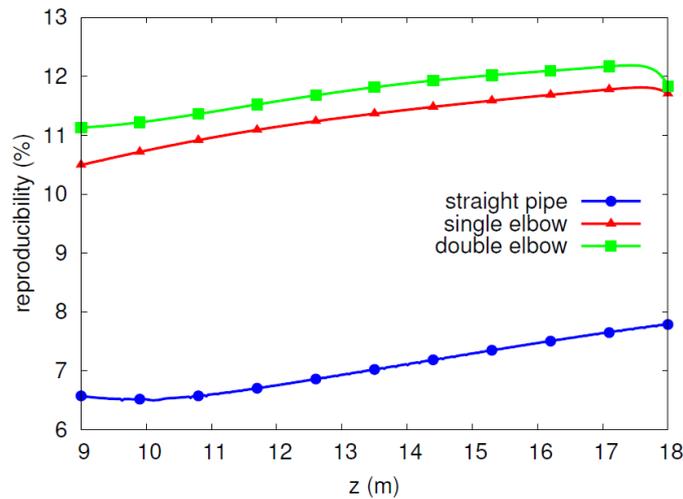

**Fig. 12.** Reproducibility of flow profiles $R$ calculated by comparing the normalised flow profile for 3 m/s with the normalised flow profile for 30 m/s at given height for various supply pipes.

## 4. Errors of the flow measurement with the S-type Pitot tube

*4.1 The measurement grid*

The Pitot tube is used to measure flow velocities in a grid of points as described in section 2.1. The minimal number of the measurement lines of the grid prescribed in [4] and in the older standard [5] is two. The minimal number of points in one line for the circular stack of the diameter 1.5 m which we consider here is 4 according to [4] and 6 according to [5]. In the following analysis we use 6 points in one line and therefore the measurement grid is exactly as in Figure 1. The measurement lines are parallel with the $x$ and $y$ axes (see Figure 4) and we consider the measurement port positions such that the vector $\vec{t}$ of the Pitot tube points in the negative sense of the $x$-axis (port above the T-junction) and the positive sense of the $y$-axis (port 90° clockwise from the T-junction when viewed from the stack outlet). Results for the height of the sampling plane in the range of 9-18 m above the stack bottom or 6-15 m above the upper part of the T-junction are investigated. The standard [4] recommends to select a measurement plane "in a section of a duct with at least five hydraulic diameters of straight duct upstream of the sampling plane and two hydraulic diameters downstream (five hydraulic diameters from the top of a stack)". In our case the hydraulic diameter is 1.5 m so the measurement plane should be at least 7.5 m above the upper part of the T-junction and at least 7.5 m below the stack outlet. These requirements are satisfied only by one plane with the height of 10.5 m above the stack bottom.

*4.2 The flow rate error $E_z$*

In this section we show results for the error (7), i.e. the error of flow rate which is caused only by the fact that the measurement grid has a finite density and therefore it gives only an approximation of the exact flow rate formula which is given by an integral of $v_z$ over the stack cross section. The error of velocity measurement in the grid points do not enter the error $E_z$. The flow rate $Q_{Mz}$ entering the formula (7) is given by the formula (1) where $v_{z\alpha}$ are directly the computed axial velocity components in the grid points. The real flow rate $Q$ is given simply as $Q = v.A$ where $v$ is the inlet velocity and $A$ is the area of the stack cross section.

Figure 13 shows how this error depends on the height of the sampling plane in the stack for various configurations of the supply pipe and various inlet velocities.

In case of the supply pipes with single and double elbow which generate a single swirl in the stack, we observe a periodic behaviour. This behaviour is caused by a velocity profile in the stack that rotates together with the swirl (see Figures 7 and 8). The velocity profile of $v_z$ in the stack cross section is not axially symmetric and therefore the error $E_z$ depends on the orientation of the sampling lines with respect to the velocity profile. Since the profile is turned with changing height by the swirl also the error changes with the height and approximately repeats when the profile is turned by 90° since the sampling grid has a 90° rotation symmetry. The periodicity is not exact since the velocity profile not only turns but also changes its shape approaching the fully developed profile.

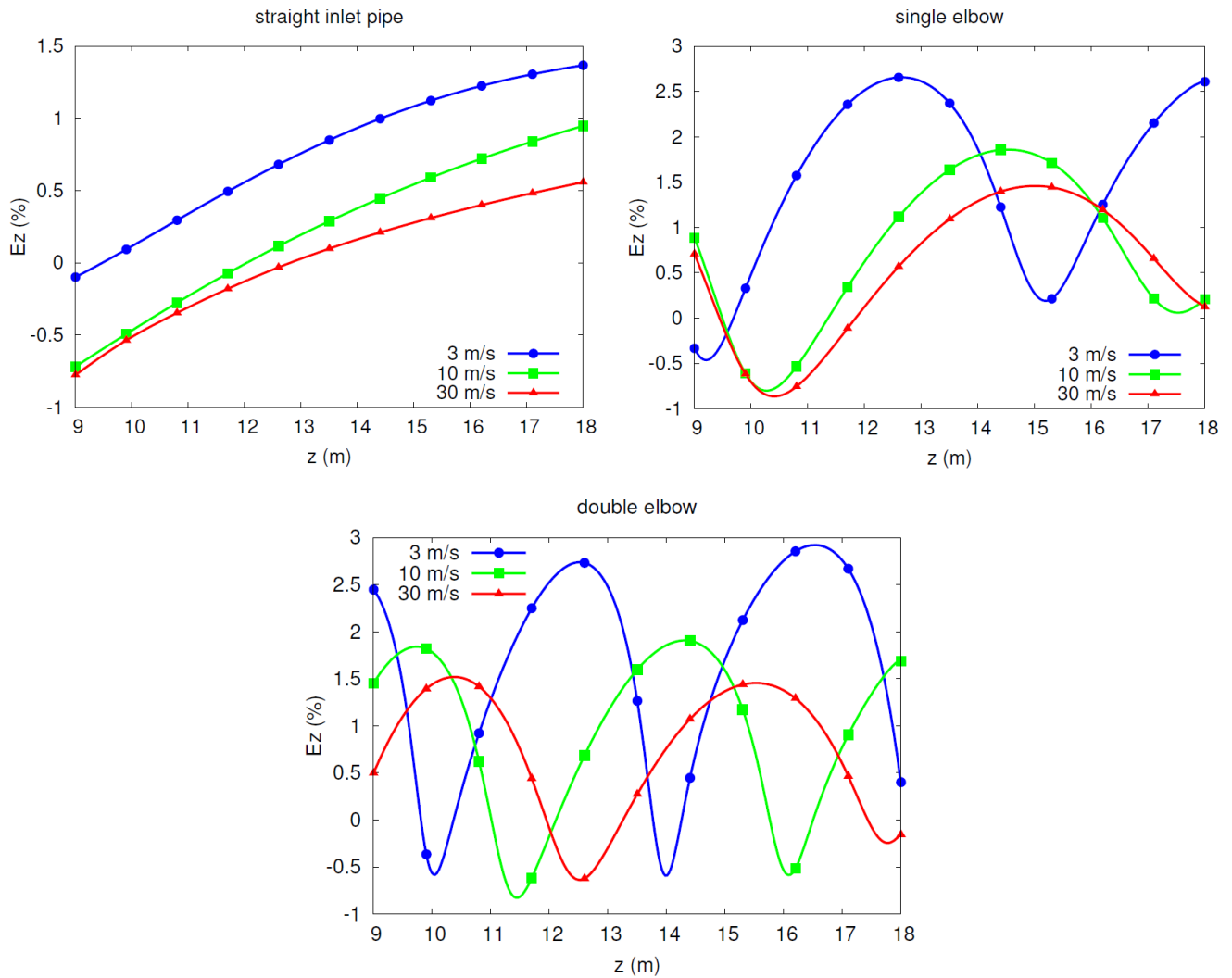

**Fig. 13.** The error of flow rate due to the finite density of the measurement grid and its dependence on the height of the sampling plane in the stack.

From Figure 13 we see that selecting the right height or orientation of the sampling lines can improve the flow measurement error by several percent. On the other hand, in practise it can be difficult to determine the optimal height or orientation of the sampling lines since the complete velocity profile and the real flow rate are usually not known, and their best knowledge comes from the reference method itself. CFD can provide certain additional insights for determining the optimal setup, but one should be aware that obtaining a reliable result with CFD also needs precise input information and the CFD results are biased by various kinds of errors (see e.g. Annex G of [8]). The CFD can at least be a useful tool for estimating the uncertainty of the reference method due to the positioning of the sampling grid.

*4.3 The flow rate error $E_i$*

Next we calculate the error (8), i.e. the error caused by both approximate integration (finite grid density) and biased velocity measurement in the grid points due to the neglected swirl. We calculate the error for both cases – when the correction described in section 9.3.5 of [3] is or is not applied (see the section 2.3).

When the correction is not applied the flow rate $Q_{Mi}$ is calculated according to the formula (1) where the velocities $v_{z\alpha}$ are the computed axial velocity components modified by the error $E(\beta, \gamma)$ given by the formula (5), which is determined from the yaw and pitch angles in the grid points, i.e. the velocities $v_{z\alpha}$ are the velocity values that would be measured by the S-type Pitot tube when the swirl is ignored.

When the correction is applied the velocities $v_{z\alpha}$ are calculated in the same way, i.e. biased with error $E(\beta, \gamma)$, in the grid points with yaw angle satisfying $|\beta| \leq 15°$ and in the grid points with $|\beta| > 15°$ the velocities $v_{z\alpha}$ are computed by modifying the axial velocity components by error $E(0, \gamma)$, i.e. the yaw angle influence to the error is eliminated by turning the S-type Pitot tube to the swirl direction.

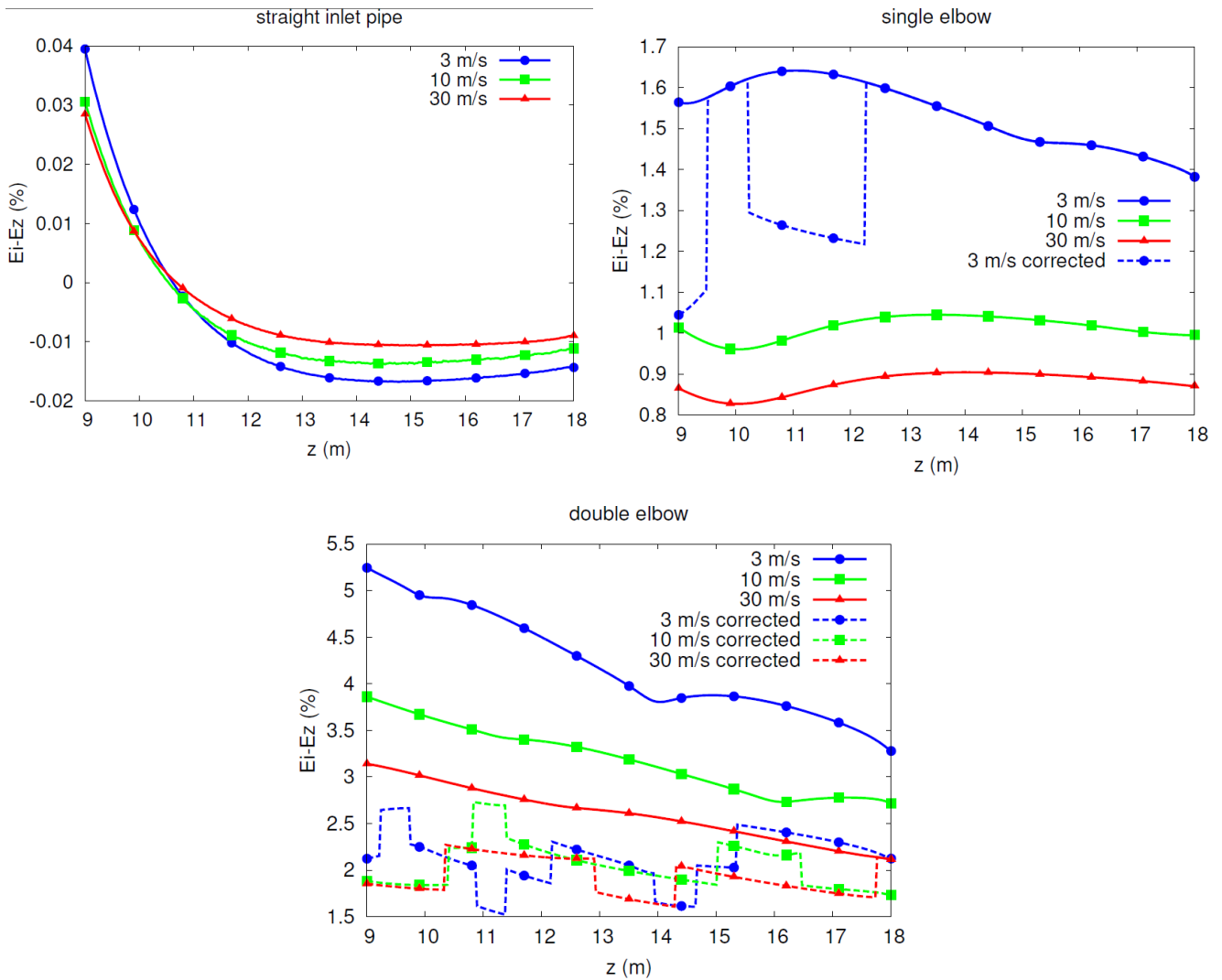

**Fig. 14.** The additional error of flow rate given by error of the velocity measurement due to neglecting the swirl in the stack, and its dependence on the height of the sampling plane in the stack.

Page 14

Figure 14 shows the difference $E_i - E_z$ and its dependence on the height in the stack for the investigated geometries, and velocities for the corrected and uncorrected cases. The yaw angles exceeding 15° occur only for the single elbow supply pipe with inlet velocity of 3 m/s and for double elbow supply pipe (all the considered velocities). Since the yaw angles evolve with height and the correction is applied when it exceeds 15°, the curves of height dependence of the error contain step changes. Each step corresponds to a height where at certain grid point the yaw angle achieves the value of 15°. For the single elbow with 3 m/s 0-1 corrected points occur, for the double elbow with 3 m/s it is 2-5 points, for the double elbow with 10 m/s it is 1-3 points, and for the double elbow with 30 m/s it is 0-2 points.

We see that the additional error due to ignoring the swirl in velocity measurements is not important for the case with straight supply pipe and it is relatively small also for the single elbow case. For the double elbow case the error becomes significant, however, it improves after applying the correction. After applying the correction it is still between 1.5 % and 2.5 %. Obviously the error can be further improved by applying the correction also for points when the yaw angle is lower than 15°, which is not forbidden by the standard [3].

## 5. Errors of the ultrasonic flow measurement

Table 3 of [8] provides informative guidance for selection of the type of automated measurement system based on the results of the pre-investigation, where the values of reproducibility, crest factor and skewness have been determined. In our models the reproducibility is always above 5 %, the crest factor is below 1.3 except a short segment of the stack with the straight supply pipe and the skewness can be below or above 1.2 depending on position in the stack. Table 3 recommends to use one cross-duct monitoring path for *cr* < 1.3 and *sk* < 1.2 and the situation with *cr* < 1.3 and *sk* > 1.2 is not considered in the table. For *cr* > 1.3 and *sk* > 1.2 two cross-duct monitoring paths (the primary and secondary paths) are recommended. The primary and secondary monitoring paths in circular ducts are defined in Annex C of [8]: "The primary monitoring path, P, shall be the path in which the maximum velocity is expected to be found. That is, in a straight line through the centre of the duct, and lying in the plane defined by the centreline of the duct being monitored and the centreline of the inlet upstream of the monitoring point." The definition of the secondary path depends on presence of an asymmetric swirl in the stack: "If swirl is not dominant asymmetric, and increased accuracy is required, it is recommended that two monitoring paths be used, parallel to P and to each other, and spaced symmetrically at 0.3 of the diameter from the centre of the duct." And: "The point in the duct where the maximum velocity is located rotates if swirl is dominant asymmetric, and consequently a monitoring path which includes the point where the maximum velocity is found cannot be determined. In this case, the secondary monitoring path, S, shall be a straight line through the centre of the duct, lying in a plane perpendicular to P."

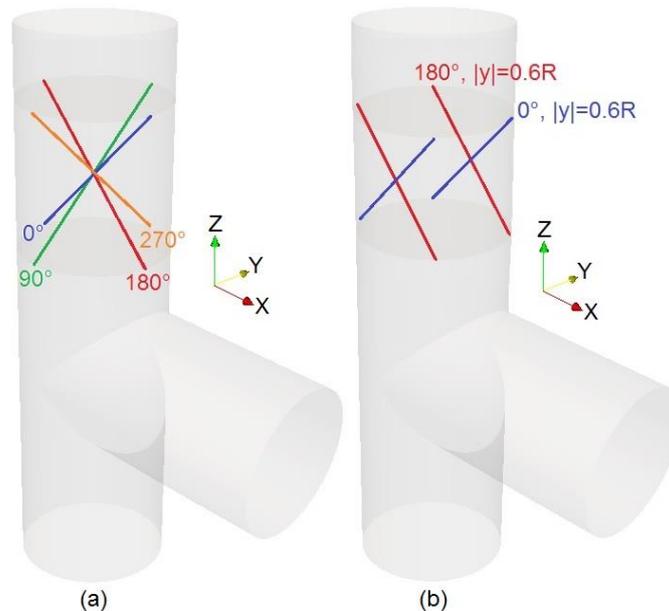

**Fig. 15.** Considered orientations of the ultrasonic beams

In this paper we consider ultrasonic beams inclined 45° to the horizontal plane as shown in Figure 15. For the straight supply pipe there is the velocity maximum at the angular position of 180° constantly (see Figure 8), therefore the primary path can be the 0° path or the 180° path in Figure 15a. Since the velocity maximum is not rotating the recommended two beam configuration corresponds to the one without dominant asymmetric swirl as shown in Figure 15b. There are two options for the parallel beams – coloured red and blue in the Figure 15b (colours in online version). In our analysis we compare all four single beam configurations from the Figure 15a and both two beam configurations from the Figure 15b.

The results for the error of the flow rate (12) as a function height of installation of the ultrasonic transducers in the stack (height of the beam centre above the stack bottom) are shown in Figure 16. The first graph of the figure is for the single beam and the second graph is for the two beam configurations. The errors are calculated without using a correction factor for fully developed profile. This factor can slightly improve the error values, however, it does not affect the comparison of various configurations. In the two beam configurations the indications from the particular beams are averaged with the same weights.

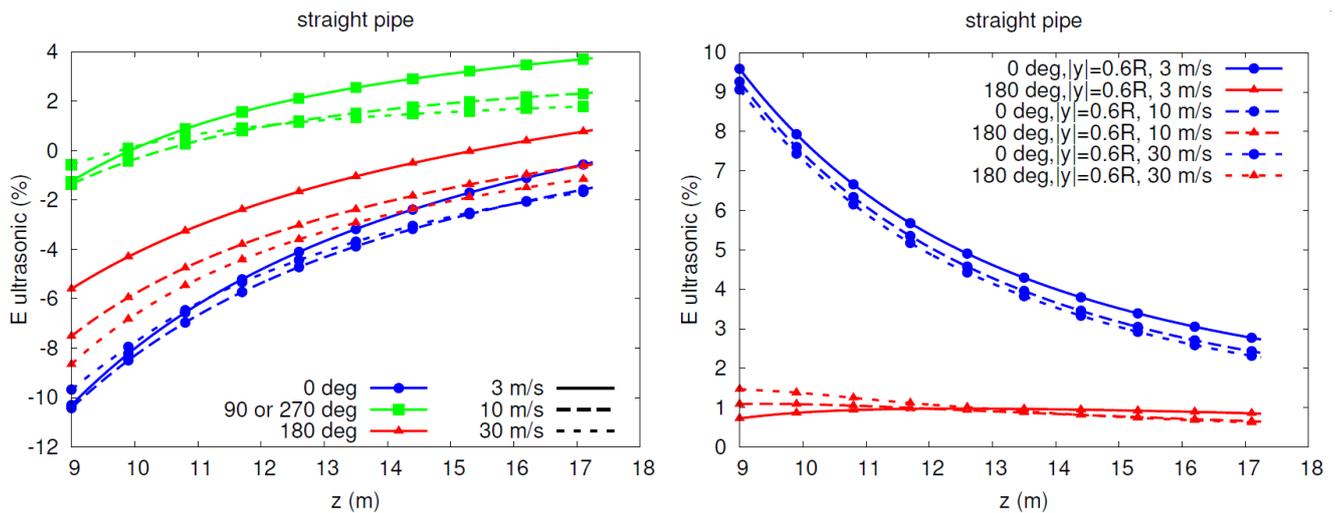

**Fig. 16.** The error of ultrasonic flow meter for the straight supply pipe and various inlet velocities as a function of height in the stack. Graphs for various one and two beam configurations according to Figure 15 are included.

We see that for the single beam configurations the choice of optimal orientation depends on the height in the stack. In lower parts the 90° orientation gives better performance (lower error) than the primary paths (as defined in [8]) with 0° or 180° orientation. In the upper part of the stack where the skewness is below 1.2 the primary paths start to give comparable or better results than the 90° orientation. For the two beam configurations we see that the right choice of the beams orientation is crucial and the 180° orientation gives significantly better result than the 0° orientation.

For supply pipes with single and double elbows the dominant asymmetric swirl is present and the velocity maximum is rotating. In these cases we investigate the four single beam configurations as shown in Figure 15a and two beam configurations that are given by pairs of the paths from the Figure 15a. We include also a result for four beam configuration given by averaging of all four paths in the Figure 15a.

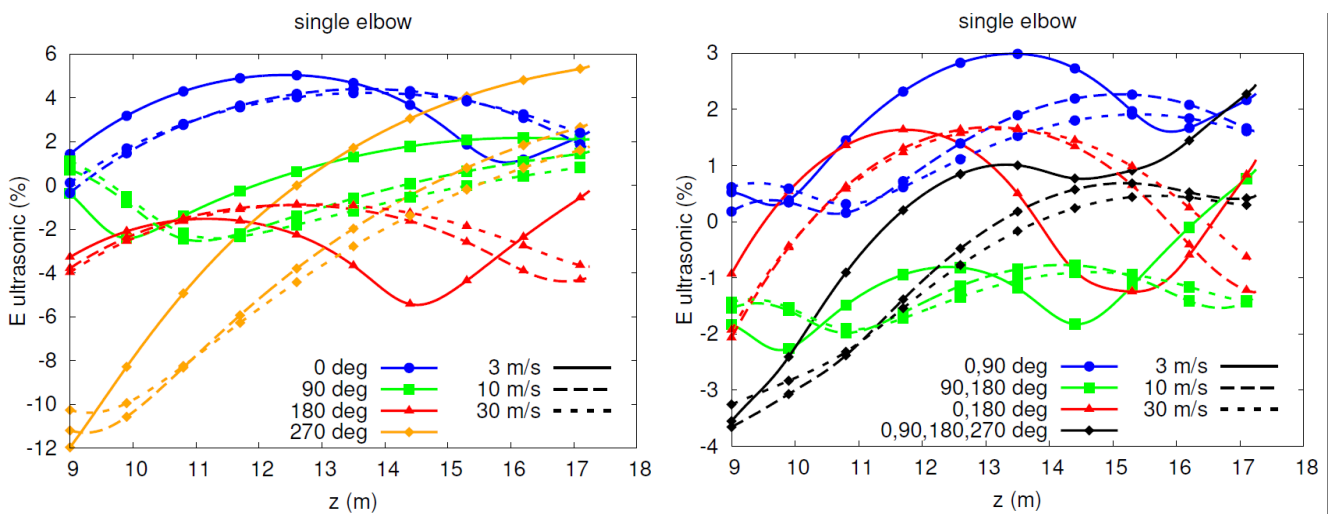

**Fig. 17.** The error of ultrasonic flow meter for the single elbow supply pipe and various inlet velocities as a function of height in the stack. Graphs for various single beam configurations according to Figure 15a are in the first figure; two beam and four beam configurations combining the paths from Figure 15a are in the second figure.

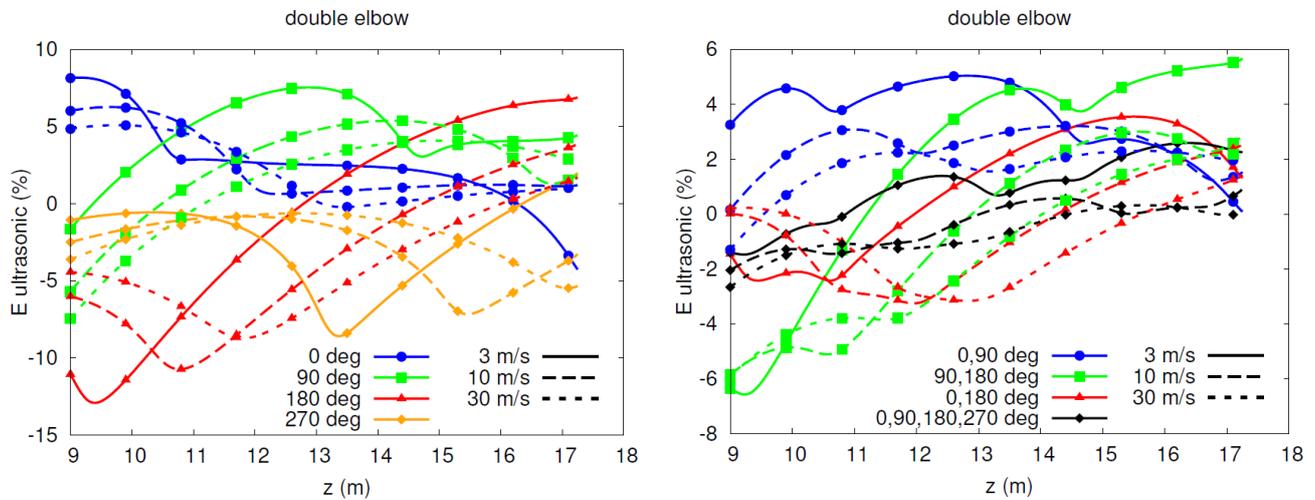

**Fig. 18.** The error of ultrasonic flow meter for the double elbow supply pipe and various inlet velocities as a function of height in the stack. Graphs for various single beam configurations according to Figure 15a are in the first figure; two beam and four beam configurations combining the paths from Figure 15a are in the second figure.

The results for the error of the flow rate (12) for the single elbow geometry are shown in Figure 17 and for the double elbow geometry in Figure 18. Again no correction factor for velocity profile is used and the flow rate for the multi path configurations is calculated by averaging the paths with the same weight. The curves for two path configurations with 270° path are not included explicitly and they can be obtained by mirroring the curve for a complementary pair across the four path configuration curve (e.g. 0°-270° curve can be obtained from the fact that the four path average is also the average of the 0°-270° curve and the 90°-180° curve).

If the pre-investigation of flow in a stack is done by CFD we can ask what configuration of paths gives the lowest error in the widest flow range for the particular height of the flow meter installation. Or if it is possible to select the height of the flow meter the model can give a hint what is the optimal height. If we are not confident that the height dependencies predicted by CFD are accurate enough we may look what configurations give the lowest error in the widest range of heights.

In Table 2 we summarise averages of errors of ultrasonic flow measurement for various configurations. The values in the table are calculated as average of absolute value of the error over height in the stack (in the section between 9 m and 18 m) and over the 3 tested flow rates. This value can be used as a criterion for the selection of the flow meter configuration if we look for the lowest error in the range of flow rates and heights tested. We see that in the case of the straight supply pipe the configuration with the lowest average error is the two path configuration with 180° orientation of the beams. For the case with single elbow the two path configuration with 0°-180° orientation gives the best result but also the single beam with 90° orientation is almost equivalent. For the double elbow case the best two path configuration is the one with 0°-270° orientation. The best single path configuration is about two times worse.

|  |  | 1 path | | | | 2 paths | | | | | | | | 4 paths |
|---|---|---|---|---|---|---|---|---|---|---|---|---|---|---|
|  | path position | 0° | 90° | 180° | 270° | 0° 0.6R | 180° 0.6R | 0° 90° | 0° 180° | 0° 270° | 90° 180° | 90° 270° | 180° 270° | 0° 90° 180° 270° |
| E (%) | straight pipe | 4.5 | 1.5 | 2.9 | 1.5 | 4.8 | 0.9 |  |  |  |  |  |  |  |
|  | single elbow | 3.3 | 1.2 | 2.3 | 4.4 |  |  | 1.5 | 1.0 | 2.3 | 1.3 | 2.8 | 2.7 | 1.2 |
|  | double elbow | 2.5 | 3.8 | 5.0 | 2.5 |  |  | 2.6 | 1.7 | 1.3 | 3.1 | 1.4 | 3.2 | 1.0 |

**Table 2** Averages of errors of ultrasonic flow measurement over height and flow rates

In practise the SRM [3] is used to calibrate the AMS in accordance with EN 14181 [10]. So the error of the ultrasonic flow meter used as AMS is determined by comparing to SRM as a reference. Therefore the AMS uncertainty is dominated by the uncertainty of the SRM.

## 6. Conclusion

Two flow rate measurement methods for stacks have been investigated for their errors in the presence of cyclonic flow by means of CFD modelling - namely the method using velocity measurements with Pitot tubes in a grid of points which is used as the standard reference method according to EN ISO 16911-1, and the ultrasonic method, which is often used as the automated measurement system in stacks according to EN ISO 16911-2. The CFD software OpenFoam was used to compute velocity fields in stacks with diameter of 1.5 m considering three different configurations of the supply pipes connected to a stack generating various swirl patterns – straight supply pipe, pipe with single 90° elbow and pipe with two out of plane 90° elbows. Three inlet velocities of the gas have been investigated – 3 m/s, 10 m/s and 30 m/s.

Two kinds of contributions to error of the standard reference method for flow measurement in stacks have been investigated. One contribution is due to the finite density of the grid, where the flow velocities are measured. In case of asymmetric velocity profile, this error depends on mutual orientation of the grid and the velocity profile and if swirl is present the maximum of the velocity profile rotates with height so the error oscillates with the height of the sampling plane in the stack. The highest oscillation in the stack configurations considered in this paper was observed for the double elbow case and inlet velocity of 3 m/s where the error is changing between approx. -0.5 % and 3 %, depending on height of the sampling plane. For real stacks where the velocity field is unknown this kind of error contributes to the uncertainty of flow rate measurement and we can see that this uncertainty contribution can be significant compared e.g. to the maximal permissible uncertainty of 2.5 % in annual emission given in [2].

Another source of error considered in this paper is the error of velocity measurement by the S-type Pitot tubes due to the inclined gas velocity vector in presence of non-axial velocity components (swirl, radial flow). The tangential inclinations above 15° (yaw angle) are corrected according to [3] but still there can be the radial inclinations (pitch angle) and also the tangential inclinations below 15° play a role. The maximal contribution to this kind of error in the stack configurations considered in this paper occurs for the double elbow case and its value is around 2.5 % (after the correction according to [3]) for all the considered flow rates. Again we see that this value is significant from the point of view of [2].

The ultrasonic flow meters used as the automated measurement system in stacks are calibrated comparing to the standard reference method. However, the installation of the ultrasonic flow meter can be optimised by pre-investigation of velocity profiles in a stack according to [8] by means of measurement or by CFD modelling. In this paper we demonstrated the flow pre-investigation by CFD and we evaluated the crest factor, skewness and reproducibility of the velocity profiles for the selected stack configurations. On top of that we evaluated the measurement error of the ultrasonic flow meters for various numbers, orientations and installation heights of the ultrasonic beams and we showed that in the stack configurations considered in this paper the correct selection of the flow meter installation can improve the measurement error by up to 10 %.

## Acknowledgement

The research has received funding from the European Metrology Research Programme (EMRP/EMPIR) under ENV60 IMPRESS and 16ENV08 IMPRESS 2. The EMRP is jointly funded by the EMRP participating countries within Euramet and the European Union. The EMPIR is co-financed by the Participating States and from the European Union's Horizon 2020 research and innovation programme.

## Appendix – Validation of the CFD model

*A.1 Brief introduction to the turbulence modelling*

There are several approaches for treating the turbulence in numerical models of fluid flow. When the turbulent flow is resolved fully by direct solution of the Navier-Stokes equations at all spatial and temporal scales (so called DNS – direct numerical simulation) the solution is very demanding in terms of computer capacity. Therefore the turbulence models are introduced treating turbulence in a simplified way and lowering the computational cost of the modelling. One class of the turbulence models called RANS (Reynolds averaged Navier-Stokes equations) is based on averaging of fields and separating the averaged fields from the fluctuation part. Another approach is based on filtering the turbulence at the lowest scale and directly computing only the larger turbulence structures (LES – large eddy simulation).

In this paper we focus on the RANS turbulence models mainly because they are a good starting point since their computational cost is the lowest. The biggest challenge when using RANS to simulate the flow in a stack is the accuracy of modelling of the turbulence. The steady RANS momentum equation can be written as:

$$\rho \left( \overline{U}_l \frac{\partial \overline{U}_k}{\partial x_l} \right) = -\frac{\partial}{\partial x_l} \left[ \bar{p} \delta_{kl} - \mu \left( \frac{\partial \overline{U}_k}{\partial x_l} + \frac{\partial \overline{U}_l}{\partial x_k} \right) + \rho \overline{u_k u_l} \right] \quad (A.1)$$

where $\overline{U}_k, \bar{p}, \rho, \mu$ are the average velocity, pressure, density and dynamic viscosity respectively, and $\overline{u_k u_l}$ is the unknown Reynolds-stress tensor, with $u_k$ being the turbulent fluctuating part of the velocity component [20]. The three components of the momentum equation together with a continuity equation are a set of four equations in ten unknowns (three velocity components, one pressure component, and six Reynolds-stress components) and additional equations must be introduced to close the system. The different equations used to close the system give rise to different turbulence models. The vast majority of turbulence models makes the so-called Boussinesq approximation where the Reynolds-stress tensor is re-written as

$$\rho \overline{u_k u_l} = \frac{1}{3} \rho \overline{u_m^2} \delta_{kl} + \rho \left( \overline{u_k u_l} - \frac{1}{3} \overline{u_m^2} \delta_{kl} \right) \quad (A.2)$$

and the second element on the right-hand side is assumed to behave as an additional shear stress such that

$$\rho \left( \overline{u_k u_l} - \frac{1}{3} \overline{u_m^2} \delta_{kl} \right) = -\mu_{turb} \left( \frac{\partial \overline{U}_k}{\partial x_l} + \frac{\partial \overline{U}_l}{\partial x_k} \right) \quad (A.3)$$

where, $\mu_{turb}$ is the turbulent viscosity to be determined to close the equations and is an isotropic scalar quantity, i.e. independent of flow direction.
This is a reasonable assumption whenever the flow is mainly isotropic but this is not the case for swirling pipe flow, where the dormient components of the Reynolds-stress tensor are activated by the presence of the bends. The eddy viscosity turbulence models may incur in problems when the flow is anisotropic. Depending on the level of accuracy required, using these turbulence models may not be sufficient anymore. This is why a thorough validation of the CFD must be carried out beforehand to ensure the results are not misleading. There is a vast body of literature dealing with these problems, e.g. [21, 22].
The obvious solution would be using an anisotropic turbulence model or employing LES, which ensure much better accuracy at the expense of much longer computational times [23, 24].

*A.2 Turbulence model selection and validation of the CFD model*

Several RANS type turbulence models were tested and compared with experimental data in physical situation similar to the one described in section 3 in order to find the model best matching reality. The experimental data were obtained from the reference [19] where a flow of air inside a T-shape channel is measured by PIV. The setup of the experiment in [19] is not exactly the same as the setups defined in section 3, but the main features of the flow through a T-shape pipe are present. The geometry of the channel is in Figure A.1. It has rectangular cross section with dimensions 20 cm x 40 cm and consists of two straight parts with length 3.3 m (supply pipe) and 6.9 m (stack) connected in a 90° T-junction, which starts 0.5 m above the stack bottom. The velocity of air at the inlet to the supply pipe is 9.85 m/s. The corresponding Reynolds number is 175 000, which is slightly below the smallest Reynolds number of 300 000 appearing in the physical situations considered in section 3 of this paper. In [19] the velocity field was measured along the lines shown in Figure A.1 in red and green colour (colours in online version). The red lines are in the centre of the duct, i.e. 10 cm above the bottom wall (denoted "c") and the green lines are in ¾ of the duct, i.e. 15 cm above the bottom wall (denoted "o"). The pair of lines denoted as INLET are 10 cm in front of the junction of the "supply" pipe, the pair OUTLET 1 is 10 cm behind the downstream corner of the junction and the pairs OUTLET 1,2,3,4 are located with a step of 30 cm along the "stack" pipe. Two velocity components are measured along each of the ten lines – component in direction of the main duct (longitudinal), i.e. *y*-component for the INLET lines and *x*-component for the OUTLET lines, and component tangential to the lines (transversal), i.e. *x*-component for the INLET lines and *y*-component for the OUTLET lines.
The mesh for the geometry given in Figure A.1 was created in *blockMesh* utility of OpenFOAM. Several mesh densities have been tested and the mesh giving mesh-converged solution was selected. The mesh consists of 27M cells, the wall cell thickness is 0.03 mm corresponding to $y+ = 1$ and the size of cells in the centre of the geometry is 3.5 mm.

Four RANS type turbulence models have been tested: *kEpsilon*, *kOmega SST*, *v2f* and *LRR*. The first three models do not cover anisotropy of the turbulence whereas the *LRR* model does (it computes all six components of the Reynolds stress tensor). The comparison of CFD results using these four turbulence models with experimental data is in Figure A.2. We see that at INLET lines all models give similar results with good agreement with the experiment. At OUTLET lines behind the T-junction the differences of the models from each other and from the experimental data start to grow and the further we are in the stack the larger the differences are. There is a back flow behind the T-junction that is challenging for all the RANS type models as we can also see in the graphs. At the end the *kEpsilon* and *LRR* models were evaluated as giving the best match with the experimental data and the *kEpsilon* model was selected for further computations. However, the match with the experimental data can still be improved e.g. by using the LES type turbulence models which are much more demanding for the computation capacity than the RANS models.

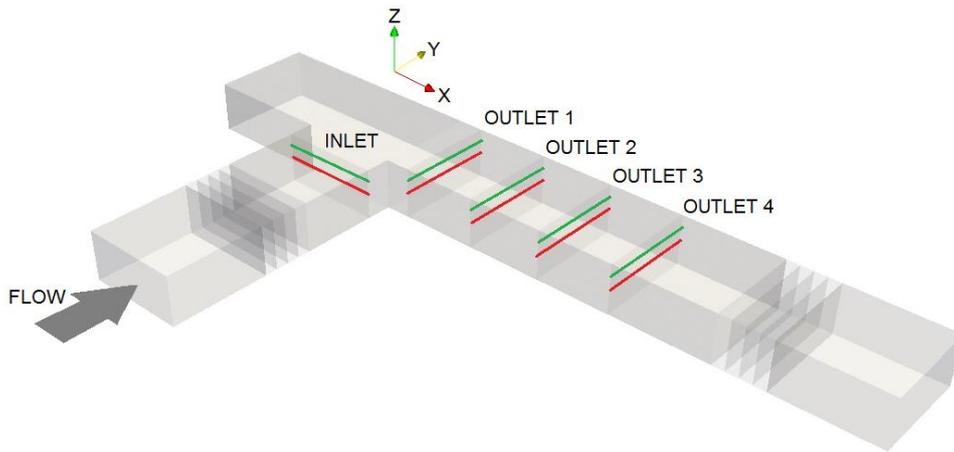

**Fig A.1.** Geometry of the duct where the experimental data used for the validation of our CFD model were obtained by PIV measurement [19].

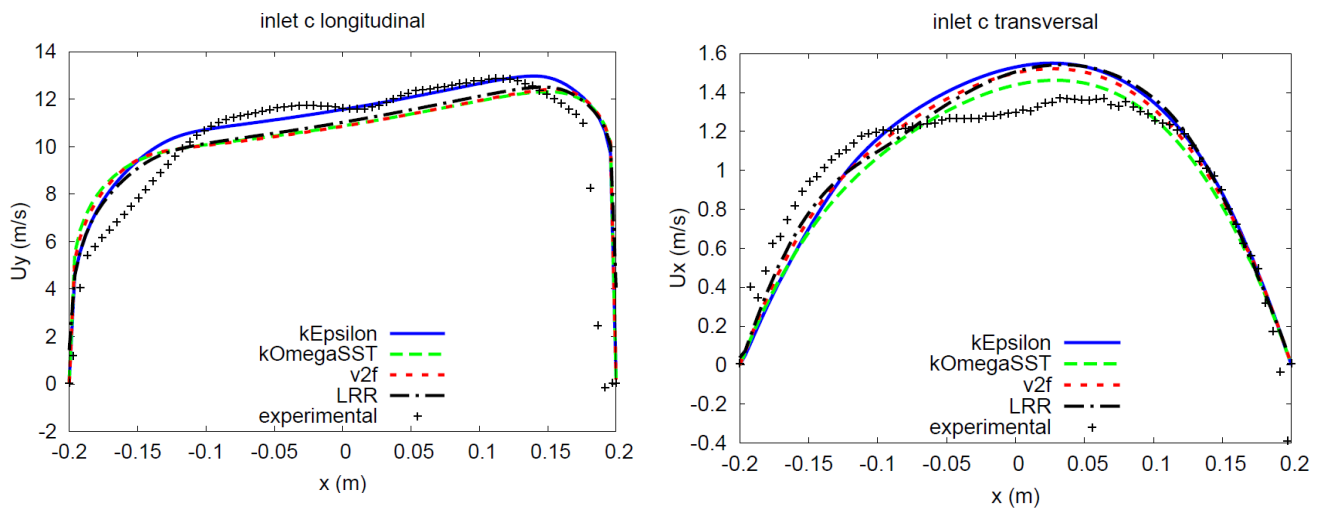

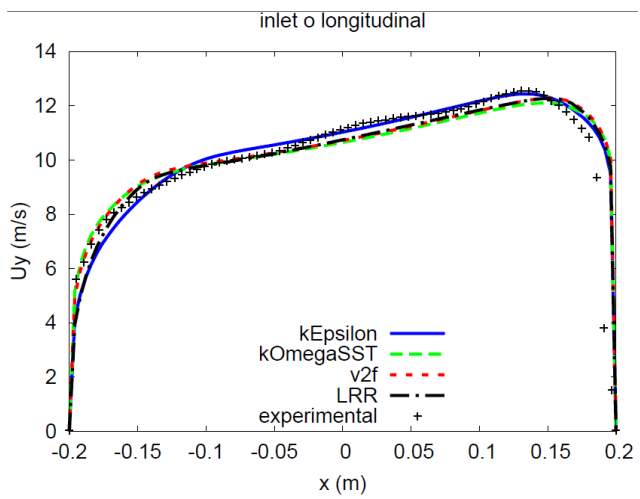
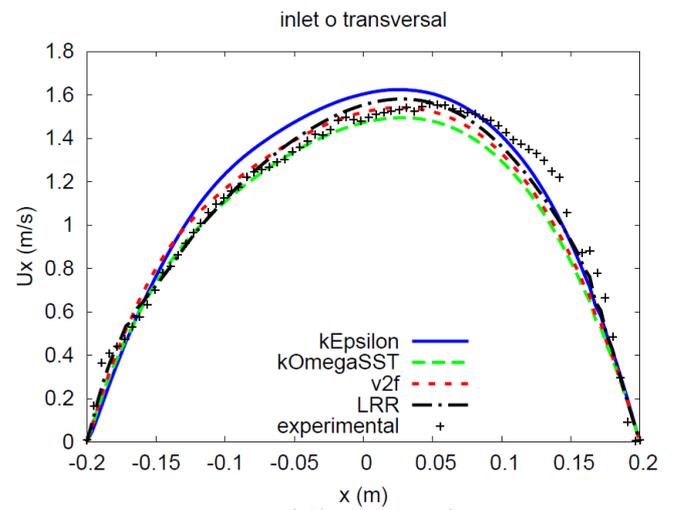
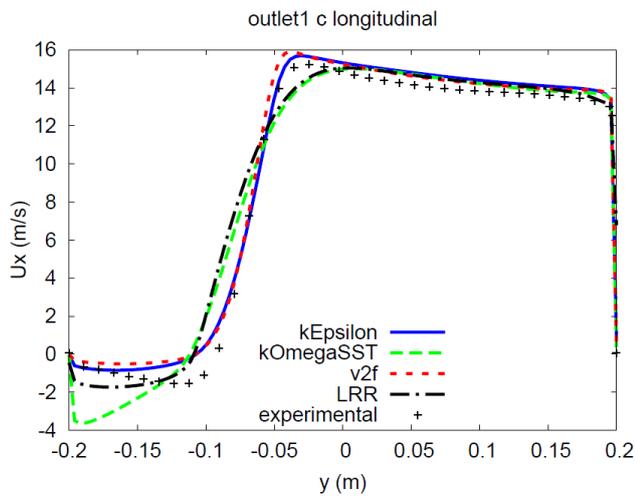
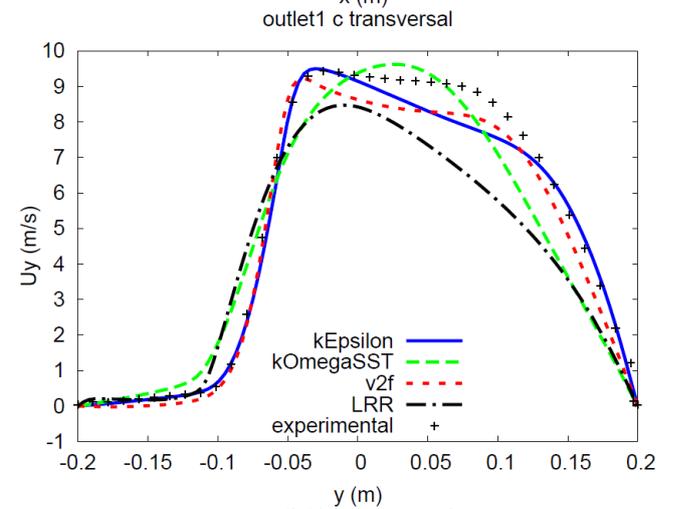
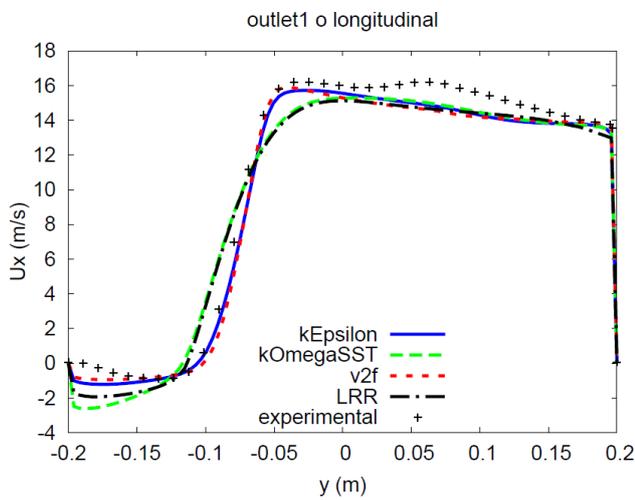
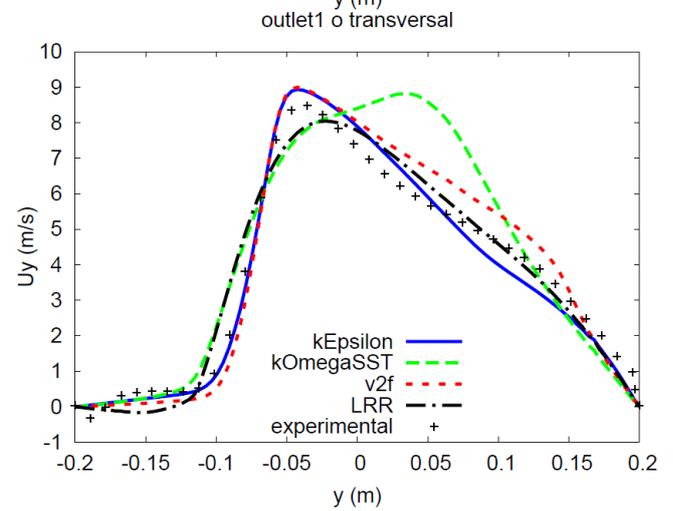

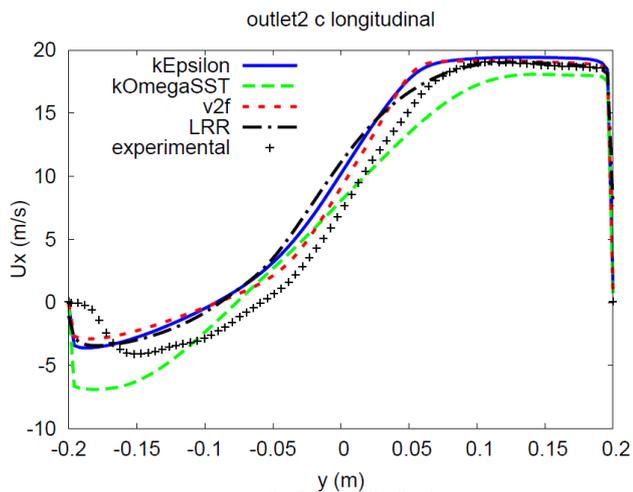
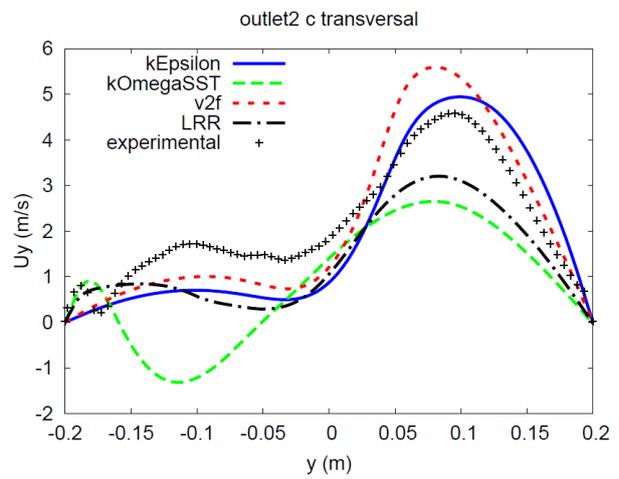
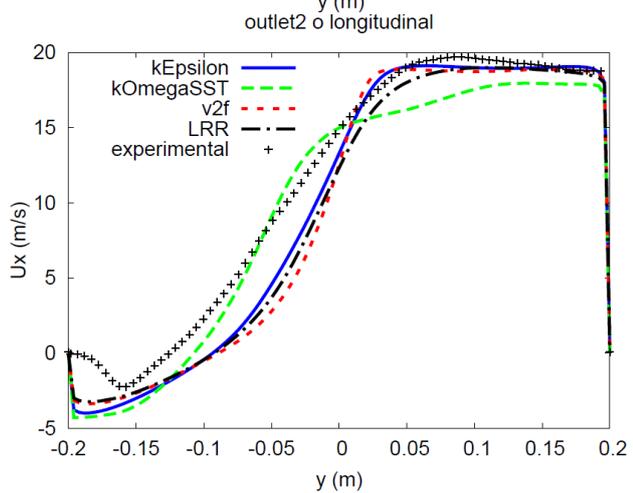
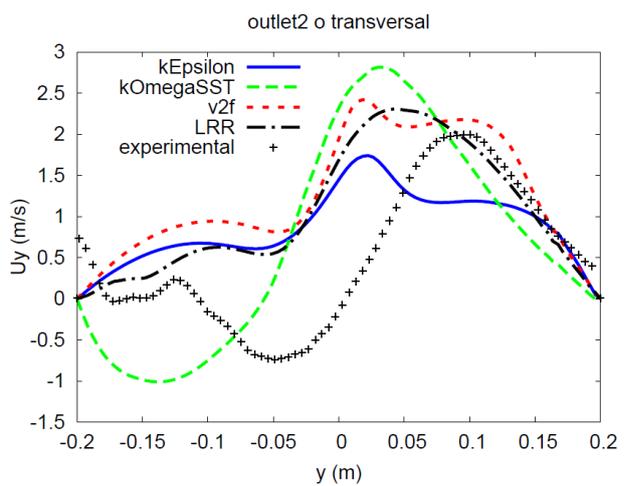
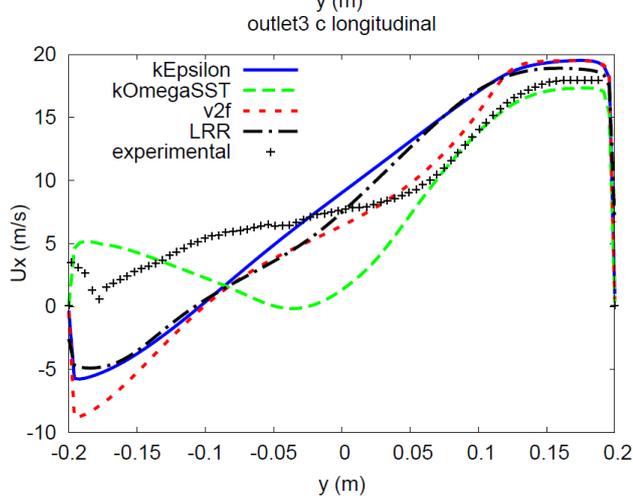
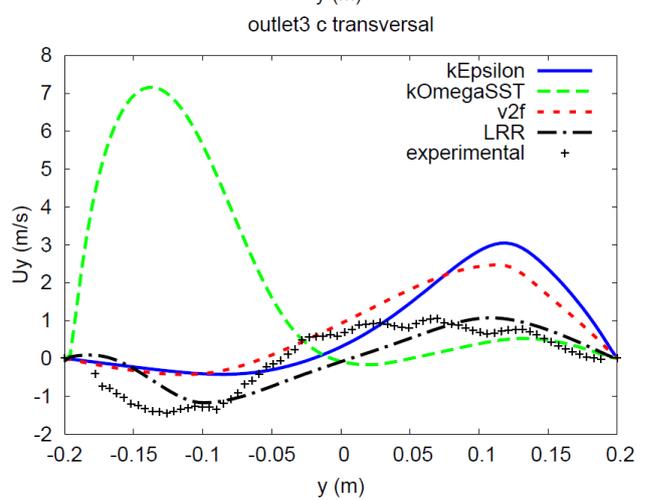

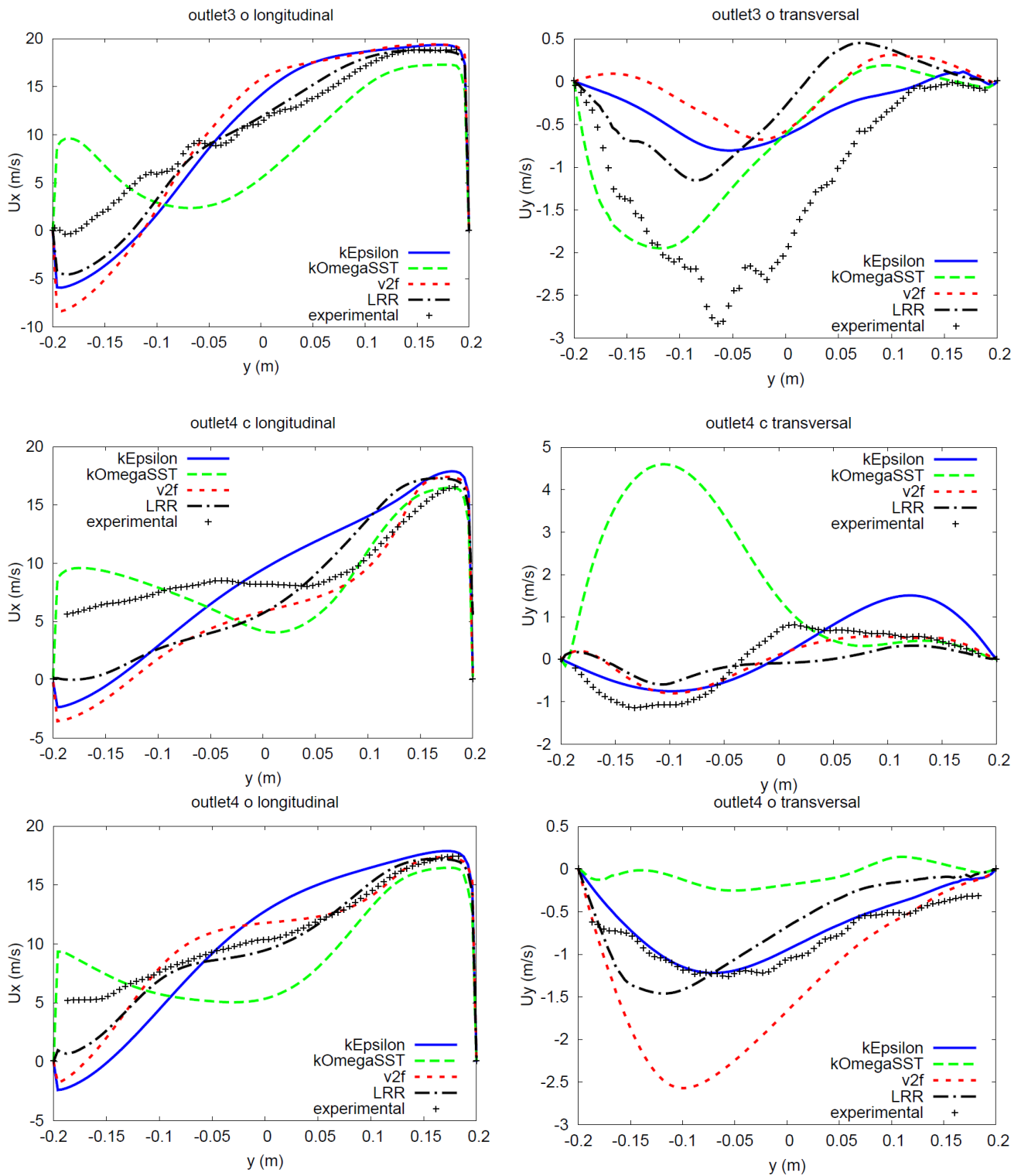

**Fig. A.2.** Data from the CFD computation with various turbulence models compared to the experimental data.

References

[1] Directive 2003/87/EC of the European parliament and of the council (2003) Establishing a scheme for greenhouse gas emission allowance trading within the Community and amending Council Directive 96/61/EC
[2] Commission Regulation (EU) No 601/2012 (2012) On the monitoring and reporting of greenhouse gas emissions pursuant to Directive 2003/87/EC of the European Parliament and of the Council